\documentclass[preprint]{aastex62}

\newcommand{\feh} {\mbox{\rm [M/H]}}
\newcommand{\fehn} {\mbox{\rm [Fe/H]}}
\newcommand{\afe} {\mbox{\rm [$\alpha$/Fe]}}
\newcommand{\mgfe} {\mbox{\rm [Mg/Fe]}}

\newcommand{\kmprs} {\mbox{\rm \,$km/s$\,}}
\newcommand{\vlos} {\mbox{\rm $V_{\rm los}$}}

\begin{document}

\title{A Hint of Three-section Halo as Seen from the APOGEE DR14}

\author{Y.Q. Chen}
\affiliation{Key Laboratory of Optical Astronomy, National Astronomical Observatories, Chinese Academy of Sciences, A20 Datun Rd, Chaoyang District, Beijing, 100101, China}
\affiliation{School of Astronomy and Space Science, University of Chinese Academy of Sciences, Beijing 100049, China}

\author{G. Zhao}
\affiliation{Key Laboratory of Optical Astronomy, National Astronomical Observatories, Chinese Academy of Sciences, A20 Datun Rd., Chaoyang District, Beijing, 100101, China}
\affiliation{School of Astronomy and Space Science, University of Chinese Academy of Sciences, Beijing 100049, China}

\author{X.X. Xue}
\affiliation{Key Laboratory of Optical Astronomy, National Astronomical Observatories, Chinese Academy of Sciences, A20 Datun Rd., Chaoyang District, Beijing, 100101, China}

\author{J.K. Zhao} 
\affiliation{Key Laboratory of Optical Astronomy, National Astronomical Observatories, Chinese Academy of Sciences, A20 Datun Rd., Chaoyang District, Beijing, 100101, China}

\author{X.L. Liang}
\affiliation{Key Laboratory of Optical Astronomy, National Astronomical Observatories, Chinese Academy of Sciences, A20 Datun Rd., Chaoyang District, Beijing, 100101, China}
\affiliation{School of Astronomy and Space Science, University of Chinese Academy of Sciences, Beijing 100049, China}

\author{Y.P. Jia}
\affiliation{Key Laboratory of Optical Astronomy, National Astronomical Observatories, Chinese Academy of Sciences, A20 Datun Rd., Chaoyang District, Beijing, 100101, China}
\affiliation{School of Astronomy and Space Science, University of Chinese Academy of Sciences, Beijing 100049, China}

\author{C.Q. Yang}
\affiliation{Key Laboratory of Optical Astronomy, National Astronomical Observatories, Chinese Academy of Sciences, A20 Datun Rd., Chaoyang District, Beijing, 100101, China}
\affiliation{School of Astronomy and Space Science, University of Chinese Academy of Sciences, Beijing 100049, China}

\begin{abstract}
Based on the $\feh$ versus $\mgfe$ diagram and distances from APOGEE data release 14, 
we compare the spatial distributions, the $l-\vlos$ diagram and the abundance gradients between
high-$\mgfe$ and low-$\mgfe$ sequences. The two sequences are clearly shown 
at $5<|Z|<10$ kpc in the metallicity range of $-1.6 <\feh <-0.7$, where the halo 
at $|Z| > 10$ kpc consists of low-$\mgfe$ stars only.
In the intermediate-metallicity range of $-1.1 <\feh <-0.7$, 
a $\mgfe$ gradient is detected for stars at $|Z|=10-30$ kpc and it
flattens out at $|Z|>30$ kpc.
The $l-\vlos$ diagram is adopted to separate halo stars from the disk
by defining the transition metallicity, which is of $\feh \sim -1.1\pm0.05$ dex
for the high-$\mgfe$ sequence and of $\feh \sim -0.7\pm0.05$ dex for the low-$\mgfe$ sequence.
The $R$ and $|Z|$ distributions for the
high-$\mgfe$ sequence, the thick disk at $-1.1<\feh<-0.7$ and the in situ
halo at $-1.6<\feh<-1.1$, have a cutoff at $R\sim15$ kpc and $|Z|\sim10$ kpc,               
beyond which low-$\mgfe$ halo stars are the main contributions.
In the metallicity range of $-1.6<\feh<-0.7$, there is a negative metallicity gradient for the high-$\mgfe$
halo at $|Z|<8-10$ kpc, while only a marginal or no slope in the $\feh$ versus $|Z|$ diagram for the low-$\mgfe$ halo
 at $|Z|<8-10$ kpc, beyond which both the high-$\mgfe$ halo and low-$\mgfe$ halo flatten out toward $|Z| > 20$ kpc.
These results indicate a complicated formation history of the Galaxy
and we may see a hint of a three-section halo, i.e. the inner in situ halo 
within $|Z|\sim8-10$ kpc, the intermediately outer dual-mode halo 
at $|Z|\sim10-30$ kpc, and
the extremely outer accreted halo with $|Z|>30$ kpc.
\end{abstract}


\section{INTRODUCTION}
It is widely accepted that our Galaxy has undergone many merging and 
accretion events \citep[e.g.][]{Bullock05}, 
and hydrodynamical simulations predict an inner halo 
(within $20-30$ kpc) dominated by relatively metal-rich stars formed within the main 
Galactic progenitor and an outer halo dominated by metal-poor stars
accreted from nearby satellite galaxies. \citep[e.g.][]{Abadi06,Zolotov09,McCarthy12}.
In this context, the
inner halo fits in line with the scenario by \citet{ELS62} that the Galaxy
formed from monolithic collapse of an isolated protogalactic cloud, while the outer
halo is constructed from many independent protogalactic fragments as proposed by \citet{SZ78}.

Observationally, two distinct halo populations, the high-$\alpha$ halo and 
the low-$\alpha$ halo, in the solar neighborhood were detected by \cite{NS10}
based on very high-precision abundances and kinematics. Stars in 
the low-$\alpha$ halo have retrograde rotation velocities and their values are
similar to $\omega$ Cen \citep{Dinescu99,Myeong18}. Thus they proposed that the 
low-$\alpha$ halo is formed from the accreted process of the Galaxy
with nearby galaxies.
With a large sample of stars within 4 kpc (which is beyond the solar neighborhood), 
\citet{Carollo10} presented evidence for dual halos: one of metal-rich stars
on mildly eccentric orbits and a second of metal-poor stars on more eccentric orbits.
They classified the former as the inner in situ halo and the latter as the
outer accreted halo. Recently, \citet{Hayes18} confirmed the
dual halos and they found that the low-$\mgfe$ population has a large
velocity dispersion with very little or no net rotation, while the high-$\mgfe$ population
shows a modest velocity dispersion around a sinusoidal variation.
Interestingly, \cite{Helmi18} made a kinematical selection of halo stars, i.e. 
total space velocities with respect to the LSR larger than 210 $\kmprs$, and 
detected a retrograde moving structure, the \textit{Gaia}-Enceladus, in the Toomre diagram.
They found that the majority of low-$\alpha$ stars within 5 kpc from the Sun are related with
the accretion of this structure, which has heated up the (thick) disk, explaining
perhaps the existence of high-$\alpha$ stars  with halolike kinematics.
Similarly, \citet{Haywood18} analyzed the blue and red 
sequences in the \textit{Gaia} HR diagram of high velocity stars in \cite{Babusiaux18} and
pointed out that 
the high-$\alpha$ stars suggested to belong to an in situ formed halo population 
may in fact be the low rotational velocity tail of the old Galactic disk heated 
by the last significant merger of a dwarf galaxy. It seems that
the formation of the Galactic halo is more complicated than theoretical predictions.

Spatially, astronomers probe the transition between the inner and outer halos
in connection with different origins of dual modes of the Galactic halo.
Interestingly, the spatial density distribution of halo stars based
on RR Lyrae stars is characterized by two power laws with 
a break radius $R \sim 45$ kpc \citep{Keller08,Akhter12}.
Theoretically, \citet{Font11} predicted that the slope of the mass density
profile decreases at large galactic radii of $R > 30$ kpc.
But many earlier works suggested the break radius within a range of $20-30$ kpc. 
Based on a spectroscopic study on dwarf stars in the SDSS survey, \cite{Carollo10} 
proposed an even smaller transition radius
at $R\sim 15-30$ kpc between the inner and outer halo. However,
\citet{Fernandez15} 
found that median abundances of $\alpha$ elements for halo stars in the SDSS survey
are fairly constant up to $R\sim$ 20 kpc, rapidly decrease between 20 and 40 kpc,
and flatten out to significantly lower values at larger distances.
In addition, \citet{Chen14} adopted red giant stars as stellar tracers and 
suggested a transition at $R\sim 35$ kpc based on the chemical and kinematical
properties of halo stars. In combining these results, it may indicate two
transition distances, at $R\sim$ 15-20 kpc and $R\sim$ 30-45 kpc in the Galactic halo.

Obviously, these transitions are inconclusive
because it is extremely difficult to obtain high-resolution spectra for stars located outside 30 kpc.
We need further studies to understand the discrepancies of the transition 
between the in situ and the accreted halo 
by using more precise data and better stellar tracers.
The current situation is that precise $\alpha$ abundances
can only be available in a very limited distance region, e.g. 
within 1 kpc of the solar neighborhood, while
for distant stars only low resolution spectra with a
limited accuracy of $\afe$ can be obtained for a large
number of stars. It is important to have precise $\alpha$ abundances 
for stars in the outer region of the Galaxy
in order to link the spatial transition of the Galactic halo 
and its origin via $\alpha$ abundance.

In this respect, the Apache Point Observatory Galactic Evolution Experiment
(hereafter APOGEE) survey \citep{Majewski17,Gunn06,Wilson10}
provides us with a good chance, because it aims to characterize the Milky Way Galaxy's 
formation and evolution through a precise, systematic, and
large-scale kinematic and chemical study. In particular, it presents abundances
for typical $\alpha$  elements, Mg, Si, Ca, and Ti, based on high-resolution
infrared spectra. Moreover, the majority of targets are giants 
so that the outer region of the Galaxy
(beyond 20 kpc) can be reached due to their high luminosities.
In the present work, we investigate the properties of the low-$\alpha$ halo and the
high-$\alpha$ halo beyond the solar neighborhood based on the APOGEE DR14 data. Specifically,
we aim to compare the $R$ and $|Z|$ distributions between the two halo populations
in order to know if low-$\alpha$ halo stars are dominating in the outer halo.
Moreover, the comparison of the metallicity gradient between the low-$\alpha$ halo
and the high-$\alpha$ halo is very important to understand their formation history.
It has been suggested that there is no clear metallicity gradient for the accreted halo
due to the merging processes of many dwarf galaxies, while the metallicity gradient
could exist for the in situ inner halo formed 
dissipatively in the inner regions and/or by disk heated stars \citep{McCarthy12}.
Finally, the behavior of the $\alpha$ abundance provides us with
key information on classifying the accreted origin for stars in the outer halo,
and thus it is interesting to investigate
the $\alpha$ abundance gradient in terms of spatial location for halo stars.

\section{The sample selection and the $\feh$ versus $\afe$ diagram}
 As part of the Sloan Digital Sky Survey IV, the APOGEE DR14 \citep{Abolfathi18,Blanton17} offers stellar parameters and
elemental abundances for 19 chemical species determined
by the APOGEE Stellar Parameters and Abundances pipeline
\citep{Garcia16}.
The sample of stars, together with stellar parameters, $\alpha$ abundances,
and stellar distances based on three methods, are taken from the APOGEE website
\footnote{http://www.sdss.org/dr14/; 
https://data.sdss.org/sas/dr14/apogee/vac/apogee-distances}.

The procedure of selecting the sample stars are as follows.
Firstly, in order to avoid the contribution from the bulge,
we exclude stars in the following two regions: (1) stars at $l < 5$ deg 
and $|b| < 10$ deg and (2) stars with $5<l<10$ deg and $|b| < 5$ deg.
Secondly, we select stars with consistent spectroscopic distances, i.e., the deviation in distance 
between the $dist_{BPG50}$ \citep{Santiago16,Queiroz18} and the $dist_{NAOC}$ \citep{Wang16}
being less than 3 kpc. 
Note that the $dist_{BPG50}$ distances were calculated using  the
measured spectroscopic parameters coupled with 2MASS photometry
via a Bayesian analysis to obtain a probability distribution
function of the distance for each star over a grid of PARSEC
stellar evolutionary models \citep{Bressan12}.
They are well consistent with {\it Hipparcos} parallaxes and
the mean error in statistical distance is 20\% according to \citet{Santiago16}.
Although \textit{Gaia} DR2 (Brown et al. 2018) provides
distances for most of the sample stars, we note that the relative error in parallax
is good (within 20\%) only for stars with $d<6$ kpc, while
for distant halo stars (e.g. $d>10$ kpc) it becomes very large (about 50\%).
Thus, we adopt the photometric and spectroscopically based distances from \citet{Santiago16}
in order
to have a consistent distance scale for the whole sample.
We have checked that these distances have no systematical deviation 
with \textit{Gaia} DR2 for common stars with relative error in parallax 
less than 10\% in \textit{Gaia} DR2.
Thirdly, we exclude stars with $R< 3$ kpc to avoid bulge stars in a more strict way
and $|Z|<0.5$ kpc to avoid too many stars from the thin disk population. 
Here $R = \sqrt{X^2+Y^2+Z^2}$ is the Galactocentric distance.
We have checked that the results are the same if we adopt 
$|Z|>0.3$ kpc or $|Z|>1.0$ kpc (to replace the $|Z|>0.5$ kpc criterion)
or without the cut of $R< 3$ kpc in the analysis.
With the main aim to probe the properties of the Galactic halo, it is expected that
the two cuts in $R$ and $|Z|$ will not affect our result because stars
at $|Z|<0.5$ kpc are mainly the thin disk and stars at $R< 3$ kpc are mainly
the bulge population \citep{Blant16}; either population has little contribution to the halo.
Finally, after applying 
a signal-to-noise cut of 50 on the APOGEE spectra and discarding
stars with unreliable parameters warned by flags
({$STAR\_BAD$, $TEFF\_BAD$, $LOGG\_BAD$, $METALS\_BAD$, $ALPHAFE\_BAD$, $CHI2\_BAD$), 61810 stars are left for further analysis.

The sample stars are giants covering the temperature range of $3600 - 5600\,K$,
the $logg$ range of $0-3.8$ dex and the metallicity range of $-2.6 - +0.5$ dex.
The $\feh$ versus $[\alpha/M]$ and  $\feh$ versus $\mgfe$ diagrams are shown 
for our sample stars (small black dots) in Fig.~1. Here, $\feh$ is the metallicity derived from the global fitting
of the APOGEE spectra and it follows the one-to-one relation with iron abundance
[Fe/H], which is derived from individual iron lines in the APOGEE DR14 catalog,
as shown in \citet{Holtzman18}.
In this paper, we adopt $\feh$ as the metallicity because \citet{Santiago16}
adopted this one to derive distances.
The typical errors are $0.040$ dex in $\feh$, $0.021$ dex in $[\alpha/M]$ and $0.030$ dex
in $\mgfe$ at $\feh \sim -0.7$ but they become large toward lower metallicity as indicated in the bottom of each panel of Fig.~1.
Clearly, there are many features in Fig.~1,
corresponding to different Galactic
populations of the Galaxy.
For comparison, we overplot the stars from  \citet{NS10}
and make linear regressions to the $\afe$ ratios, which are averaged values of four $\alpha$ elements (Mg, Si, Ca, and Ti) taken from the \citet{NS10} paper,
and to the $\mgfe$ ratio alone. The relations are of  $\afe=-0.088(\pm0.015)\fehn+0.221(\pm0.014)$ and $\mgfe=-0.041(\pm0.017)\fehn+0.300(\pm0.016)$
for the high-$\alpha$ sequence, and of
$\afe=-0.202(\pm0.023)\fehn-0.036(\pm0.026)$ and $\mgfe=-0.178(\pm0.032)\fehn-0.035(\pm0.036)$ for the low-$\alpha$ sequence.
With these regressions, the main features in the APOGEE sample
are consistent with the thick disk/high-$\alpha$ halo and the low-$\alpha$ halo
in the NS10 paper. And it is easy to identify the other two populations in the APOGEE data, i.e. the thin disk and the very metal-poor halo ($\feh < -2$), which are not investigated by NS10 paper.
Specifically, in the metal-rich part of $\feh >-0.7$, the low-$\alpha$ 
sequence is the thin disk while the high-$\alpha$ sequence is the thick disk. 
Stars in the intermediate-metallicity range of $-1.6 < \feh < -0.7$ also
show two trends, the nearly flat $\afe$ ratio around 0.3 dex
merging into the thick disk on the metal-rich end
and the decreasing $\afe$ with increasing metallicity somewhat overlapping
with the thin disk. 

In this paper, we adopt the division of high-$\alpha$ and low-$\alpha$
sequences based on $\mgfe$, rather than $[\alpha/M]$, because 
comparing a single element abundance might be more straightforward than
comparing a global ratio of $[\alpha/M]$ in
the APOGEE dataset. Moreover,
the separation between high-$\alpha$ and low-$\alpha$ sequences 
at $\feh \sim -0.7$ is larger by using $\mgfe$ than using the global $[\alpha/M]$.
The upper and lower limits of 
high-$\mgfe$ and low-$\mgfe$ populations are defined to be stars within the
dashed lines along the regressions of \citet{NS10} data shifted by
0.08 dex upward and downward. The shift of 0.08 dex
 is chosen so that all blue and red dots from \citet{NS10}
can be included within the high-$\mgfe$ and low-$\mgfe$ sequences, respectively.
With the help of these lines, we can see that
the two sequences mixed together at $\feh < -1.6$, which is classified as the
metal-poor halo in this work.

\begin{figure}
\plotone{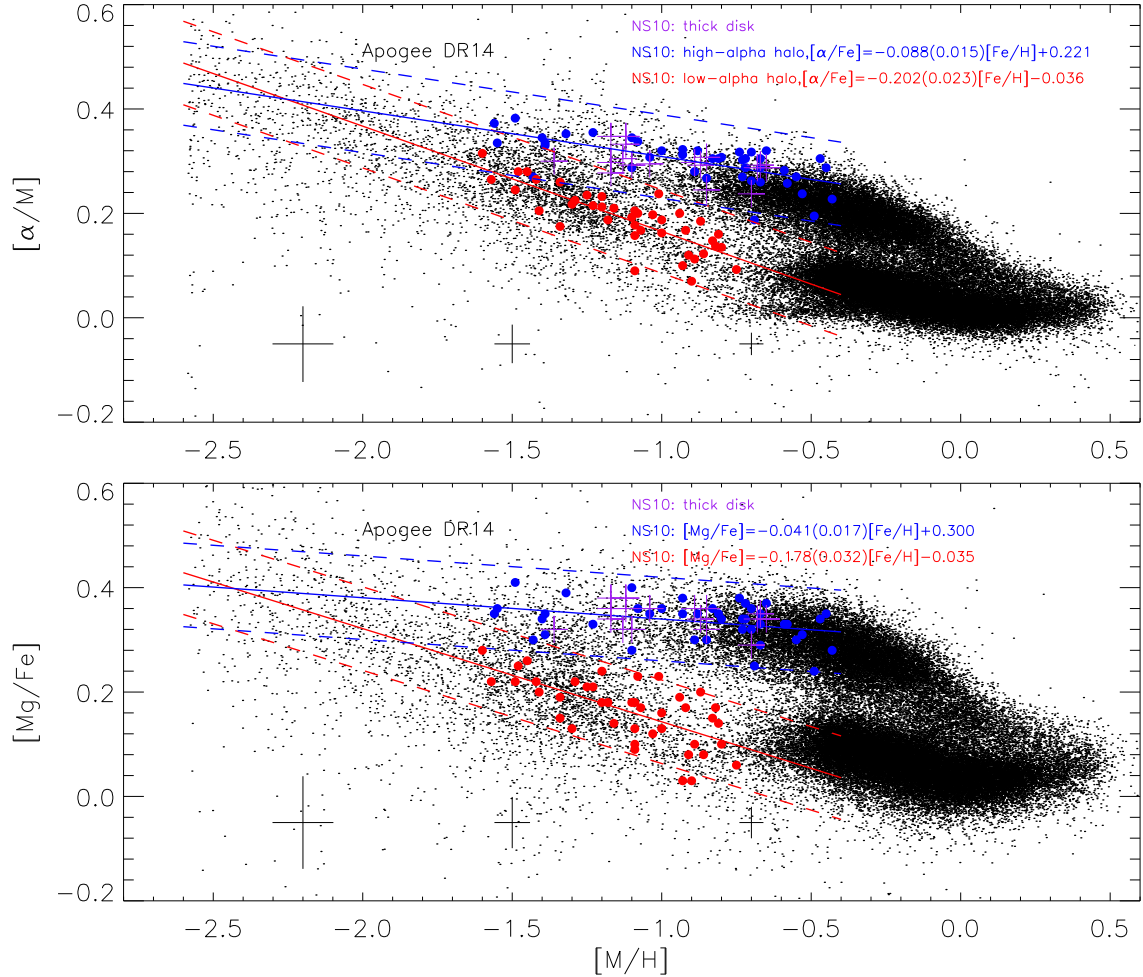}
\caption{The $\feh$ versus $[\alpha/M]$ (upper) and the $\feh$ versus
$\mgfe$ (lower) diagrams for our sample stars (black dots).
Low-$\alpha$ halo, high-$\alpha$ halo, and thick disk stars
from \citet{NS10} are shown in red dots, blue dots, and
purple pluses, respectively. Linear regressions to these data are shown 
by red and blue solid lines, and they are shifted by
0.08 dex (dash lines) to selet different groups of stars. Typical
errors at $\feh \sim -0.7,-1.5$, and $-2.2$ are indicated at the bottom.}
\label{MgFef1}
\end{figure}

\section{Halo stars with $|Z|> 5$ kpc and the $\mgfe$ gradient}
In the literature, various methods, e.g. selecting nonrotating stars
or high velocity stars, can be adopted to pick out halo stars. But the 
calculation of space velocity requires distance, radial velocity, and proper motions,
which are not available for distant stars. 
With the need of stellar distance and stellar position, the calculation 
of $|Z|$ is somewhat easier and thus it provides the simplest way 
to pick out halo stars. That is, we can select stars locating at high $|Z|$,
beyond the reach of the disk. 
Since the scale height of the thick disk is in the range 
of $0.5-1.0$ kpc \citep{Blant16}, the main population at $|Z|=1-3$ kpc 
is the thick disk. In view of this, it has been suggested that stars at $|Z|>4$ kpc 
belong to the halo according to \citet{Xue08}. In this work, we adopt a more strict
one, i.e. stars with $|Z|> 5$ kpc being the halo sample.

The halo sample is divided into five groups in a $|Z|$ interval
of 5 kpc within 20 kpc and of 10 kpc or larger
beyond 20 kpc (in order to include enough stars for statistics), i.e.
$5<|Z|<10$ kpc, $10<|Z|<15$ kpc, $15<|Z|<20$ kpc, 
$20<|Z|<30$ kpc and $|Z|>30$ kpc. The distributions of stars in the $\feh$ versus $\mgfe$ 
diagram are shown in Fig.~2 where all stars with $|Z|>0.5$ kpc are shown 
in small black dots for comparison. It shows that, at $5<|Z|<10$ kpc, there
are two (both high-$\mgfe$ and low-$\mgfe$) sequences in
the metallicity range of $-1.1 <\feh < -0.3$,
while the high-$\mgfe$ sequence with $\feh > -1.1$ nearly disappears at $|Z|> 10$ kpc.
Fig.~3 shows that the relative star number
between the high-$\mgfe$ sequence within $-1.1 <\feh < -0.6$ and the
low-$\mgfe$ sequence within $-1.6 <\feh < -1.1$ decreases with $|Z|$ and
is close to zero at $|Z|\sim 8$ kpc.
In view of this, we further pick out stars at $8<|Z|<10$ kpc,
shown as purple dots in the upper left panel of Fig.~2. It is found that most of them belong to 
the low-$\mgfe$ sequence clumping at $\feh \sim -1.4$, as already noticed by
\cite{Helmi18}, with a low fraction of high-$\mgfe$ stars belonging
to the thick disk and/or the in situ halo.
Finally, we pick out the most distant stars with $|Z|>30$ kpc,
marked by blue dots in the lower right panel. Interestingly, they
are found to have systematically lower $\mgfe$ than those at $20<|Z|<30$ kpc
at a given metallicity. 

Inspired by this, we suspect that there might be a vertical gradient in the $\mgfe$ ratio 
for the halo at $|Z|> 10$ kpc. 
Fig.~4 shows the $|Z|$ and $R$ versus $\mgfe$ diagrams
for all stars at $10<|Z|< 60$ kpc and for three separate metallicity ranges,
$-1.1 <\feh< -0.7$, $-1.6 <\feh< -1.1$, and $-2.6 <\feh< -1.6$.
The reasons for the metallicity divisions at $-0.7$ and $-1.1$ will be given in the next section; they correspond
to the transitions between the halo and the disk for the low-$\mgfe$
and high-$\mgfe$ sequences.
Meanwhile, as described in Sect. 2, the two sequences start to mix together at $\feh \sim -1.6$
and thus this is chosen as the start of the metal-poor population in
the present work.
In Fig.~4, the average $\mgfe$ at given $|Z|$ and $R$
intervals are indicated by open red squares and single linear fits to them
for the whole $|Z|$ or $R$ ranges are shown by blue dash lines. 
However, the single linear regression 
is not a good solution
for all stars at $|Z|>10$ kpc and for stars in
the metallicity range of $-1.1 <\feh< -0.7$. Moreover, it seems that there
is a flat trend of $\mgfe$ versus $|Z|$ and $R$ at distances larger
than 30 kpc for all the three metallicity ranges. Although 
single linear regression might give a reasonable fit to the data
for the metallicity ranges of $-1.6 <\feh< -1.1$ and $-2.6 <\feh< -1.6$,
we note that a significant number of stars at $|Z|< 30$ kpc have $\mgfe\sim 0.0$
and they deviate significantly from the single
linear regression. Instead, they seem to follow the flat trends
of stars at distances larger than 30 kpc as an extension into 
smaller distances of 10-30 kpc.

In view of this, we perform a segment test
to the data based on the $fit.seg$ method in R-software by taking into 
account the errors. We find a break at $|Z|=31$ kpc
for all stars and stars with $-1.1 <\feh< -0.7$, i.e. at $|Z|\sim 30$ kpc
in view of the large error at this distance.
Then the data are divided into two segments, $|Z|=10-30$ kpc
and $|Z|=30-60$ kpc, and two linear regressions
are fitted separately.
For stars with $-2.6 <\feh< -1.6$, the regressions are performed
for $|Z|< 30$ kpc, beyond which only a few stars are present and/or
the coverage of $|Z|$ is too narrow to fit. 
Clearly, in the metallicity range of $-1.1 <\feh< -0.7$,
there are two-segment $\mgfe$ gradients with different
slopes, i.e.
$\mgfe=-0.0071(\pm0.0010)|Z|+0.181(\pm0.020)$ ($10<|Z|< 30$ kpc) and
$\mgfe=+0.0051(\pm0.0034)|Z|-0.281(\pm0.137)$ ($30<|Z|< 60$ kpc),
with the coefficients of Pearson correlation of $-0.96$ and $+0.73$
and scatters of 0.011 and 0.032 dex,
respectively. When the single linear fit to the whole range is adopted,
the $\mgfe$ gradient is of $\mgfe=-0.0049(\pm0.0013)|Z|+0.131(\pm0.039)$ ($10<|Z|< 60$ kpc) with a reduced Pearson coefficient of $-0.79$ and a larger scatter of 0.046 dex. It
seems that two-segment $\mgfe$ gradients are preferred for the metallicity range of
$-1.1 <\feh< -0.7$.
The two-segment $\mgfe$ gradient is not prominent for the two
metallicity ranges of $-1.6 <\feh< -1.1$ and $-2.6 <\feh< -1.6$, partly
due to the large uncertainty in abundance at the metal poor end.
Further investigation with high precision spectroscopic data for 
more stars at the outer Galaxy is needed to clarify this topic.

\begin{figure}
\plotone{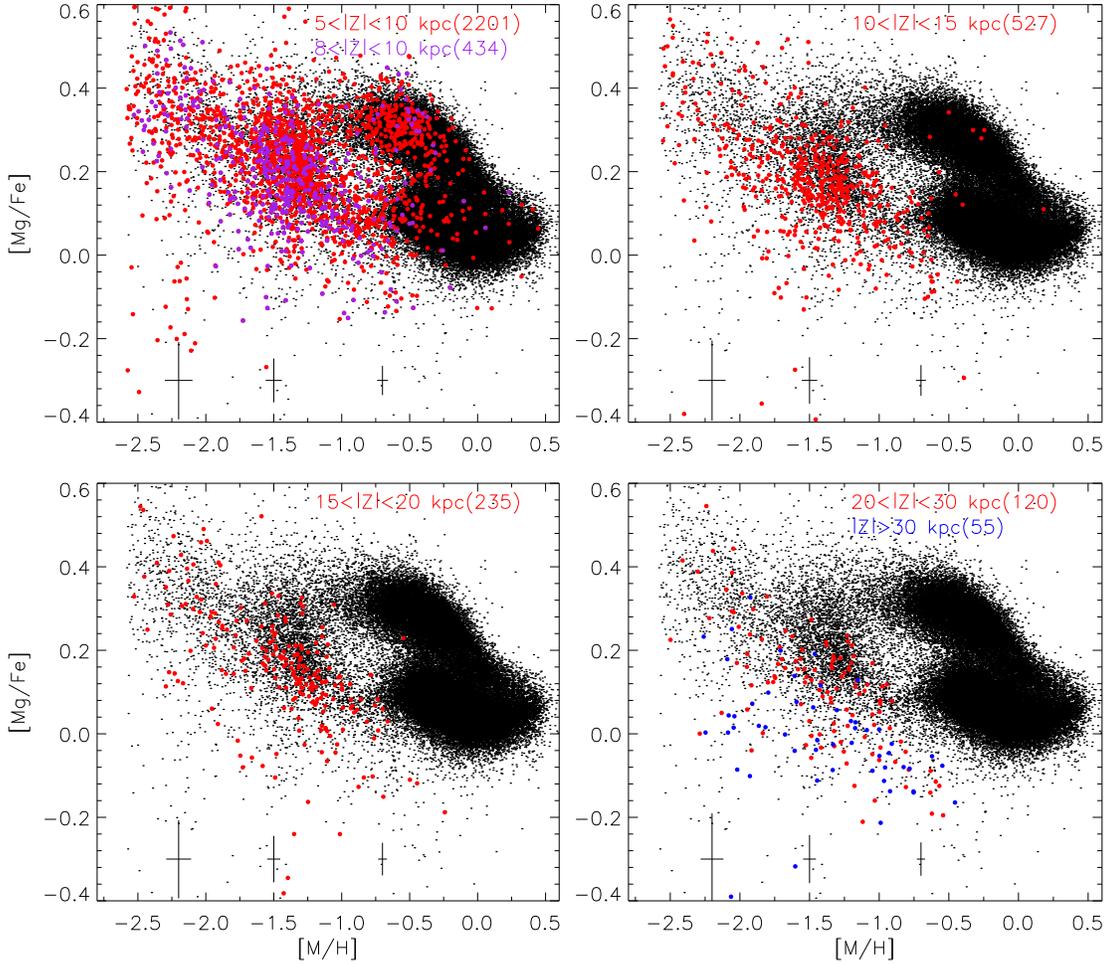}
\caption{The $\feh$ versus $\mgfe$ distributions for halo stars
at $5<|Z|<10$ kpc (including $8<|Z|<10$ kpc), $10<|Z|<15$ kpc, $15<|Z|<20$ kpc, 
$20<|Z|<30$ kpc and $|Z|>30$ kpc. The star number at each $|Z|$ interval is 
given in brackets. All stars with $|Z| > 0.5$ kpc in our sample 
are shown by small balck dots for comparison.}
\label{AMZ}
\end{figure}

\begin{figure}
\plotone{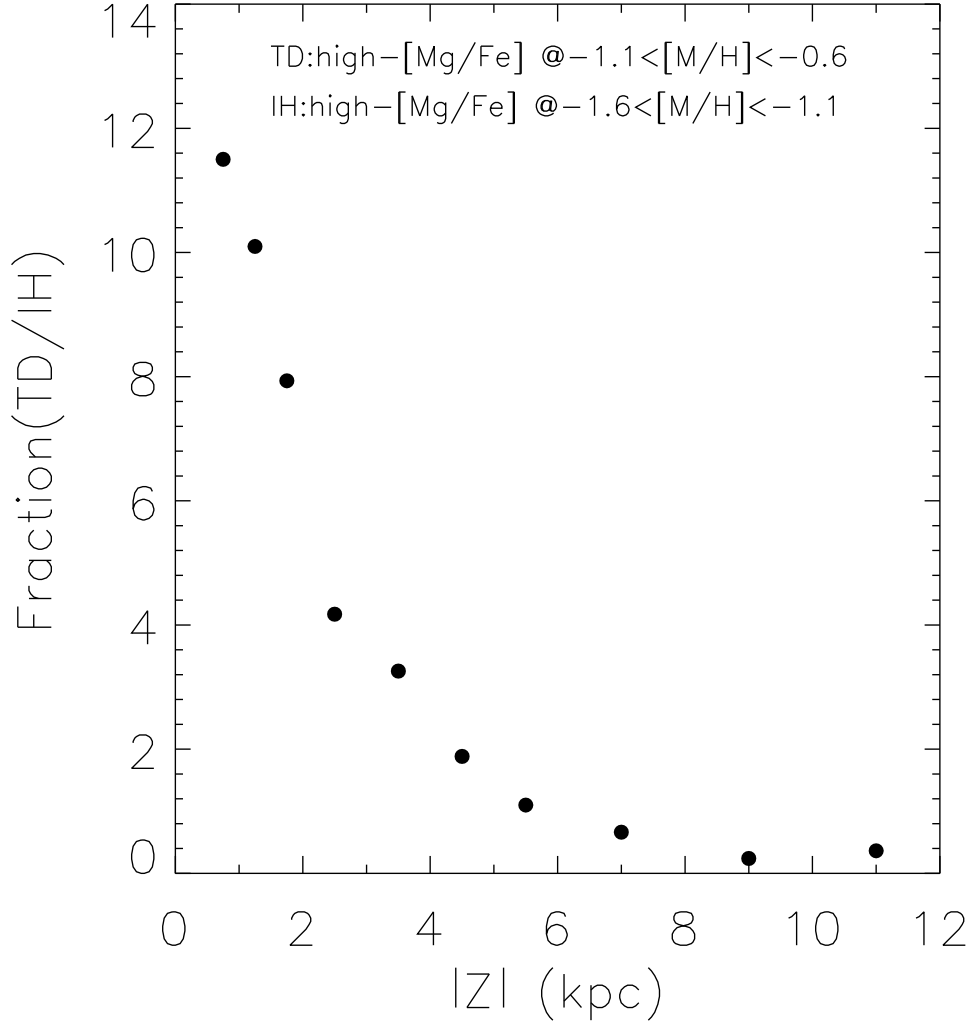}
\caption{Relative star number 
between the thick disk (TD, $-1.1 <\feh<-0.6$) and the in situ halo (IH, $-1.6 <\feh<-1.1$) as a function of
$|Z|$ in the high-$\mgfe$ sequence.}
\label{TD2IHvsZ}
\end{figure}

\begin{figure}
\plotone{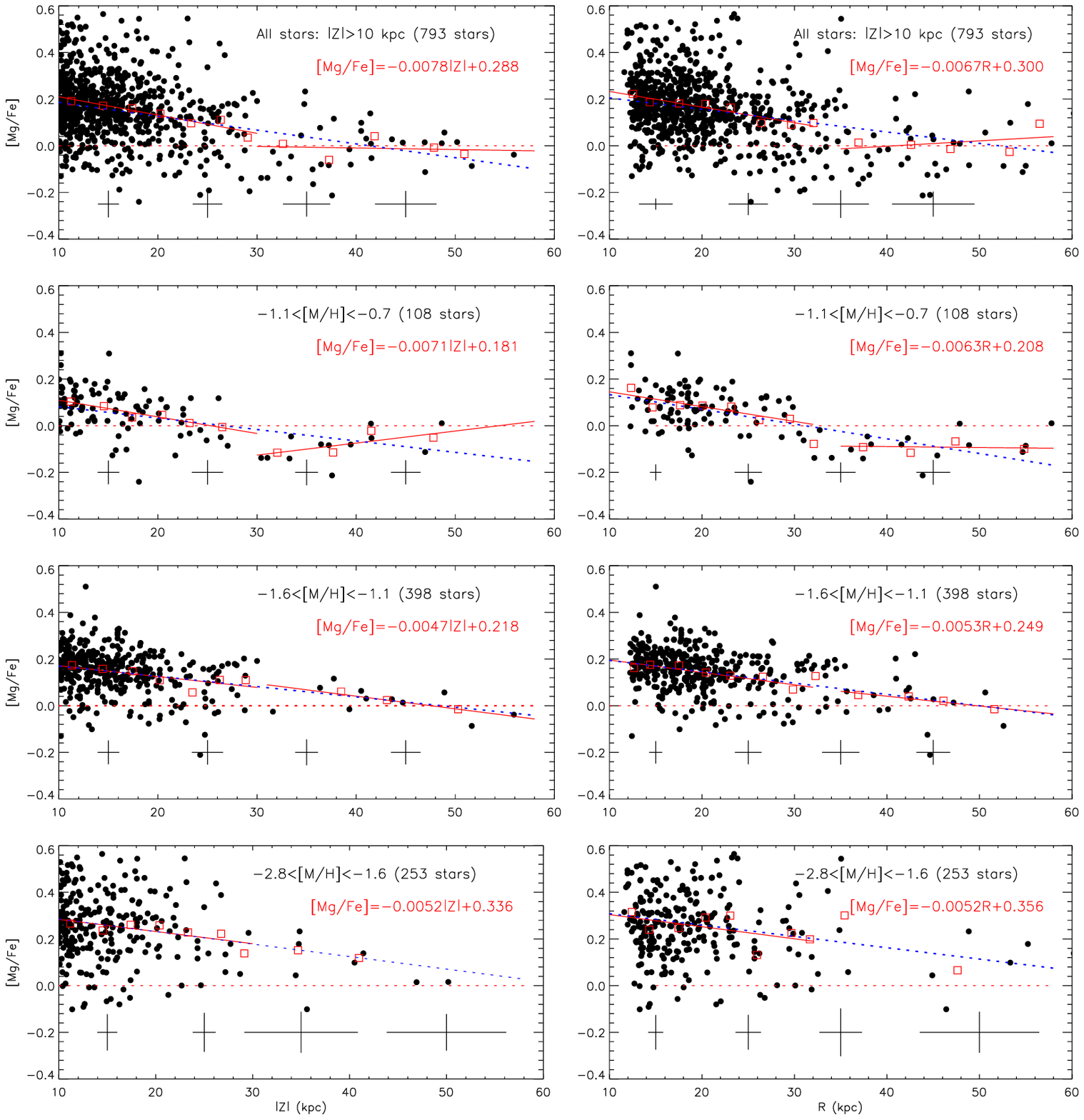}
\caption{The $\mgfe$ versus $|Z|$ (left) and $R$ (right) diagrams 
for all stars at $|Z|>10$ kpc (top) and stars with three metallicity ranges
($-1.1 <\feh<-0.7$,  $-1.6 <\feh<-1.1$ and $-2.6 <\feh<-1.6$).
Open squares indicate the average $\mgfe$ ratio at a given $|Z|$ for each bin.
Blue dashed lines indicate single linear regressions to open squares for the whole range and red solid lines for the two separate ranges divided at 30 kpc.
The star number is given in brackets and the separate regression for
the first segment (10-30 kpc) is presented in each panel.} 
\label{MgZ10}
\end{figure}

\section{Different properties between high-$\mgfe$ and low-$\mgfe$ sequences}
\subsection{The $l$ versus $\vlos$ diagram}
Although stars at $|Z|> 5$ kpc mainly belong to the halo population, there are 
more halo stars located at $|Z|< 5$ kpc. Therefore, it is important
to select halo stars in a more specific way. 
Following \citet{Hayes18}, the $l$ versus $\vlos$ diagram can be 
adopted to classify disk stars, for which $\vlos$ shows a sinusoidal
variation with Galactic longitude $l$, but there is no such trend 
for halo stars. Note that
the high-$\mgfe$ sequence includes the thick disk and the in situ halo,
and the low-$\mgfe$ sequence consists of the thin disk and the accreted halo.
{That is, both sequences have disk and halo stars.}
Thus, we investigate the $l$ versus $\vlos$ and $\sigma(\vlos)$ diagrams
for high-$\mgfe$ and low-$\mgfe$ sequences separately in Figures 5 and 6.

Stars within the two dashed red lines in Fig.~1 are selected and 
the $l$ versus $\vlos$ diagrams are plotted based on two bin sets:
the odd and the even bin sets. For the  high-$\mgfe$ sequence,
the odd bin set has $\feh$ from $-0.4\pm0.1$ to $-1.4\pm0.1$
and the even bin set from $-0.5\pm0.1$ to $-1.3\pm0.1$ in steps
of 0.2 dex. 
It is found that the sinusoidal trend in the $l$ versus $\vlos$ diagram 
starts to disappear at $-1.3<\feh<-1.1$, and there is a hint of such
a trend at $-1.2<\feh<-1.0$. Thus 
we define the transition metallicity of $\feh\sim -1.1\pm0.05$ to separate
the thick disk and the in situ halo for the high-$\mgfe$ sequence.
In a similar way, the odd bin is set to vary $\feh$ from 
$-0.4\pm0.1$ to $-1.2\pm0.1$ and the even bin set from 
$-0.3\pm0.1$ to $-1.1\pm0.1$ in steps of 0.2 dex for the low-$\mgfe$ sequence.
The transition metallicity is defined to be $\feh\sim -0.7\pm0.05$ 
because the sinusoidal trend disappears at $-0.7<\feh<-0.9$,
while there is still a sign of such a trend at $-0.8<\feh<-0.6$. 

Furthermore, the amplitude of
the sinusoidal function in each bin also varys with metallicity
for both high-$\mgfe$ and low-$\mgfe$ sequences, which is consistent
with the fact that the disk is kinematically hot and the halo
kinematically cold. Quantitatively, the amplitude in the sinusoidal 
function, calculated to be the mean $\vlos$ for stars 
at the $l \sim 85\,deg$ bin,} decreases from $82$ to $45\,\kmprs$ between the two adjacent
metallicity bins around $\feh\sim-1.1$ in the high-$\mgfe$ sequence,
and a big drop from $120.6$ to $28.5\,\kmprs$
between the two adjacent metallicity bins at $\feh\sim-0.7$ is found
in the low-$\mgfe$ sequence. Meanwhile, the dispersion
around the mean $\vlos$ as a function of $l$ vary significantly 
with metallicity. As shown in Fig.~6, they
start to become large at the transition metallicity ranges
of $-1.3<\feh<-1.1$ for the high-$\mgfe$ sequence
and of $-0.9<\feh<-0.7$ for the low-$\mgfe$ sequence, generally following in line with the two metallicity divisions via the $l$ versus $\vlos$ diagram.
These variations in amplitude and dispersion of $\vlos$ provide
further support for the separation between the disk and the halo
by using this method.
In this analysis, we include stars with $|Z| > 5$ kpc although
this technique usually works for stars not too far from the Galactic plane.
And we have checked that the transition metallicities are exactly the same 
if we limit our sample of stars at $0.5<|Z|< 5$ kpc only.
Finally, we notice that the two metallicity divisions
are consistent with the results in the literature. For the low-$\alpha$ 
sequence, \citet{Hawkins15} suggested that the low metallicity end of 
the thin disk is at $\feh \sim -0.7$, below which it belongs to the 
accreted halo, and the metallicity division at $\feh \sim -1.1$
for the high-$\mgfe$ sequence is exactly the same as
that of \citet{Fernandez18a}.

\begin{figure}
\plotone{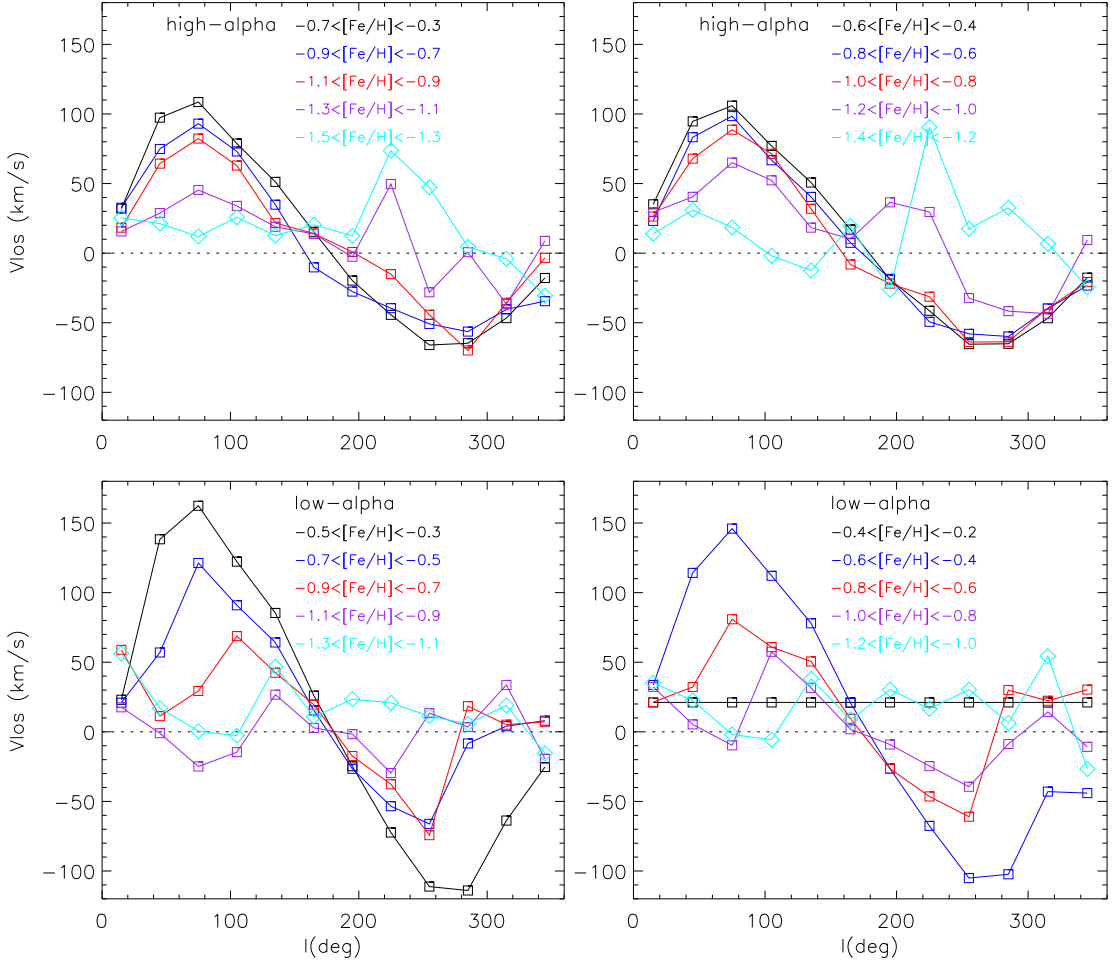}
\caption{The $l$ versus $\vlos$ diagrams for the high-$\mgfe$ (upper)
and low-$\mgfe$ (bottom) sequences with
different colors indicating stars in different metallicity intervals.}
\label{AMlb}
\end{figure}

\begin{figure}
\plotone{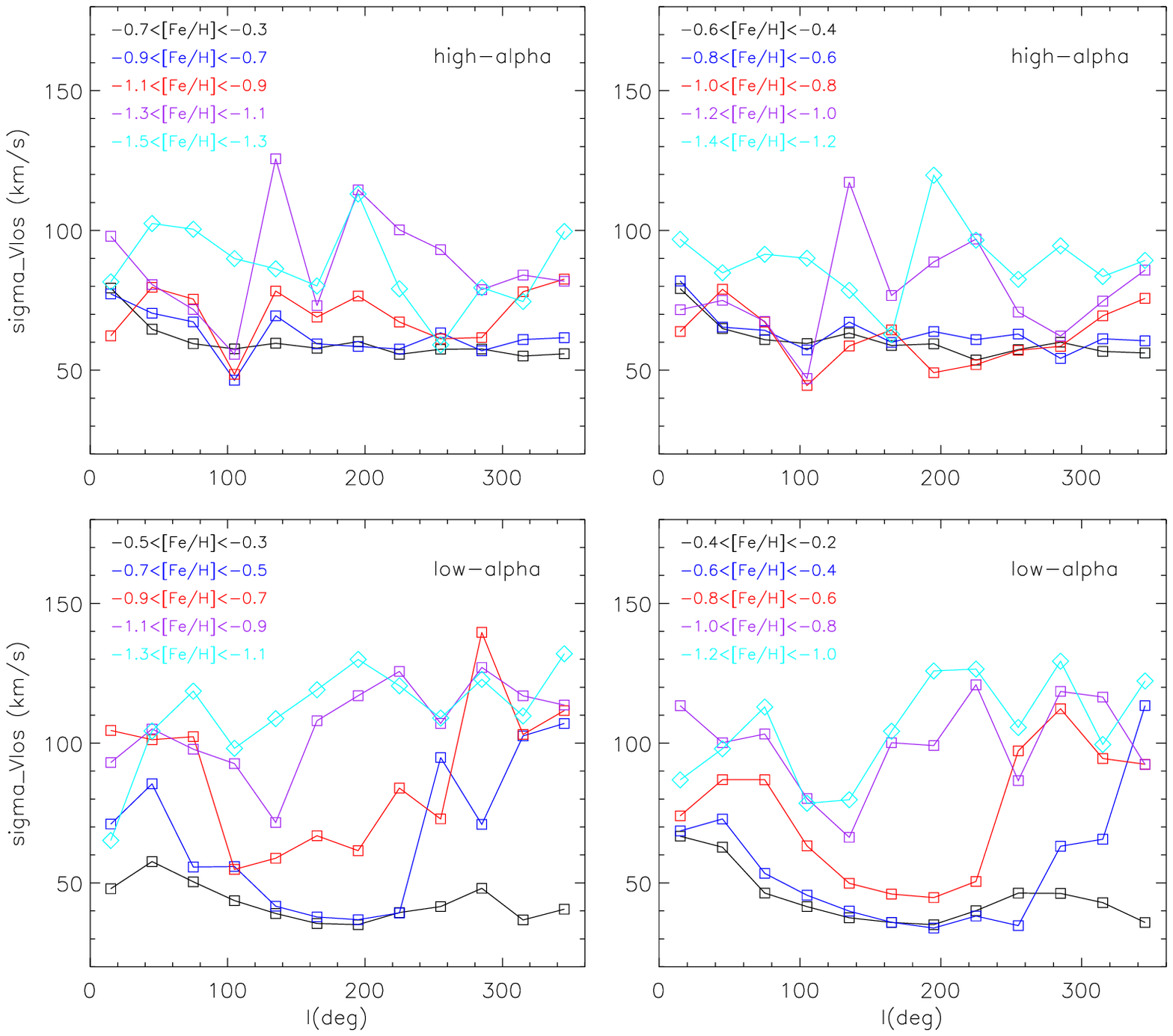}
\caption{The $l$ versus $\sigma(\vlos)$ diagrams for the high-$\mgfe$ (upper) and low-$\mgfe$ (bottom) sequences. The symbols and colors are the same as those in Figure 5.}
\label{AMlc}
\end{figure}

\subsection{The $R$ and $|Z|$ distributions}
After setting the metallicity boundaries between the halo and the disk
for the two $\mgfe$ sequences in the $\feh$ versus $\mgfe$ diagram, we divide our
sample into four groups of stars, i.e. the high-$\mgfe$ thick disk with $-1.1<\feh<-0.7$ (cyan dots), the high-$\mgfe$
in situ halo with $-1.6<\feh<-1.1$ (blue crosses), and the low-$\mgfe$ accreted halo 
at the same two metallicity ranges (red and purple dots). Then we compare
the relative distributions of star number among the four groups of stars 
as functions of $R$ (Galactocentric distance) and $|Z|$ (vertical distance to the plane)
in the left panel of Fig.~7.
It shows that they all peak at $R \sim 8-10$ kpc
and $|Z| < 5$ kpc, which indicates that the halo and the disk mix
significantly in the inner Galaxy.
However, there is a big difference in relative fraction of star number
between high-$\mgfe$ and low-$\mgfe$ sequence at larger distances. In particular,
the high-$\mgfe$ thick disk (cyan line) and the in situ halo (blue line)
 show a cutoff at $R\sim 15$ kpc and $ |Z|\sim 10$ kpc,
beyond which the low-$\mgfe$ accreted halo (red and purple lines) still contributes
substantially, as is clearly shown in the corresponding subfigures of Fig.~7. 

In the right panel of Fig.~7, the low-$\mgfe$ accreted halo at the
same two metallicity ranges (red and purple dots) are compared with
more metal-poor halo with $-2.0<\feh<-1.6$ (orange) and $-2.6<\feh<-2.0$ (aqua).
Similarly, the two groups of stars from the low-$\mgfe$ accreted halo peak at $R \sim 8-10$ kpc
and $|Z| < 5$ kpc as the reference samples, but
they persist to overpopulate at larger distances of $R> 15$ kpc and $ |Z|>10$ kpc.
Although the overpopulation of low-$\mgfe$ accreted halo
at $-1.1<\feh<-0.7$ and $-1.6<\feh<-1.1$ is weaker than the comparison with
stars in the high-$\mgfe$ sequence, red and/or purple lines
outnumber the orange/aqua lines
by $3-5$ times, as seen in the subfigures of the right panel of Fig.~7.

\begin{figure}
\plotone{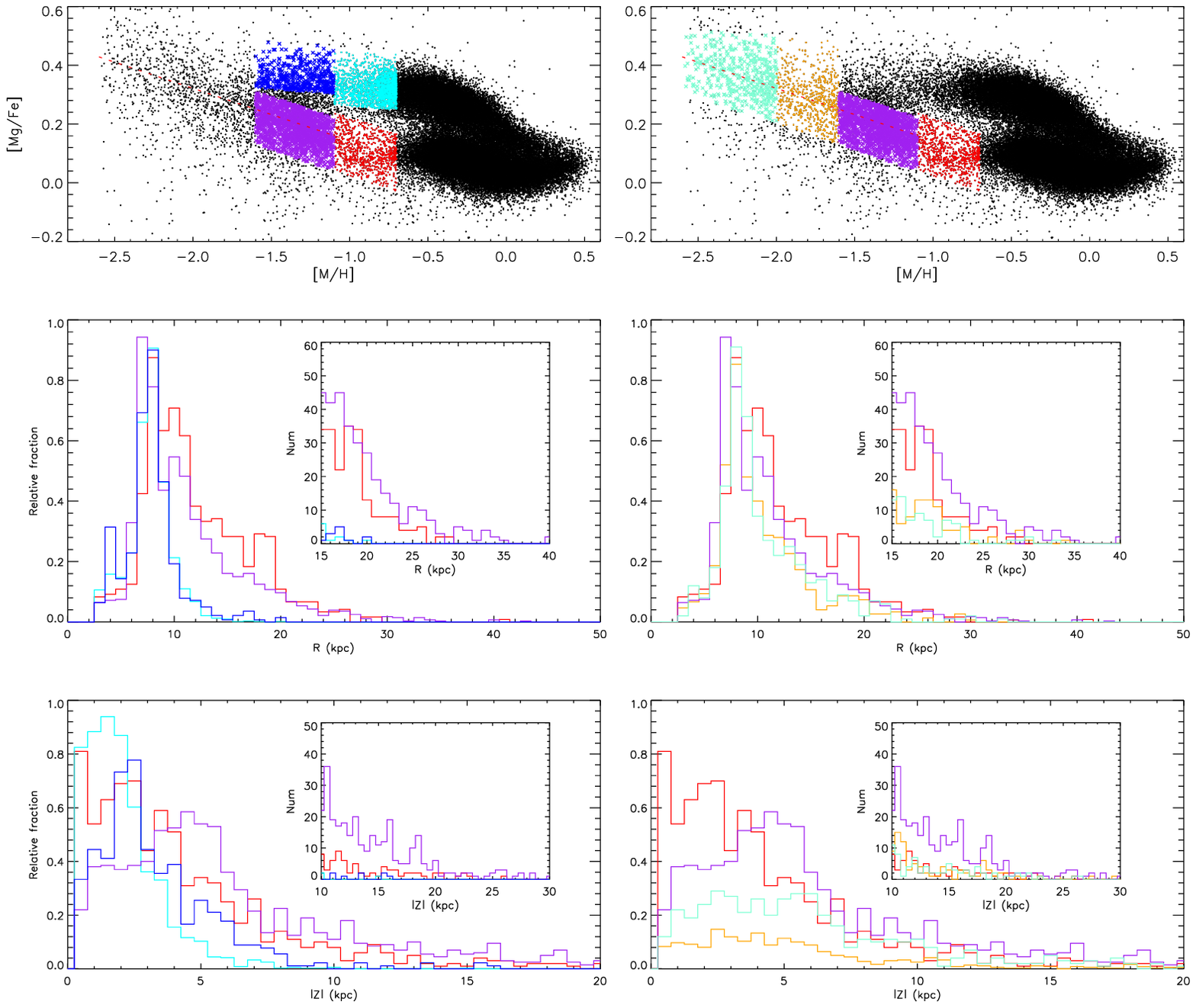}
\caption{Comparison of the low-$\mgfe$ accreted halo at $-1.1<\feh<-0.7$
(red dots and lines) and $-1.6<\feh<-1.1$ (purple dots and lines)
with two sets of reference groups. Left: the high-$\mgfe$ thick disk with $-1.1<\feh<-0.7$ 
(cyan dots and lines) and the high-$\mgfe$ in situ halo with 
$-1.6<\feh<-1.1$ (blue dots and lines). Right: the more metal-poor halo 
with mixing high-$\mgfe$ and low-$\mgfe$ populations
at $-2.0<\feh<-1.6$ (orange dots and lines) and $-2.6<\feh<-2.0$ 
(aqua dots and lines). In the subfigures, we compare the 
distributions of star number at larger distances of $R> 15$ kpc and $ |Z|> 10$ kpc.} 
\label{MgFe}
\end{figure}

\subsection{The metallicity gradients}
It is interesting to compare the metallicity gradient between the 
high-$\mgfe$ and low-$\mgfe$ sequences in order to
understand the formation scenarios of the Galactic halo.
In the left panel of Fig.~8, we plot the $\feh$ versus $|Z|$ for stars with
$-1.6<\feh<-0.7$, which is the metallicity range where
the two $\mgfe$ sequences are well separated. Interestingly, for stars
at $|Z|<8$ kpc, metallicity gradients exist for both high-$\mgfe$ and 
low-$\mgfe$ sequences with coefficients of Pearson correlation of 0.96
in both cases. However, we notice that the former has a steeper
slope ($-0.053\pm0.005$) than the latter ($-0.037\pm0.006$). 
At larger $|Z|$, mean metallicity seems to remain flat for 
the low-$\mgfe$ sequence. 
On close inspection of the bottom left panel of Fig.~8, it is
found that the main reason for this difference may come from
the inclusion of the high-$\mgfe$ thick disk stars
with $-1.1<\feh<-0.7$.
It may be more reasonable to limit this comparison at a narrow metallicity 
range of $-1.6<\feh<-1.1$. As shown in the right panel of Fig.~8,
there is no significant trend for the low-$\mgfe$ sequence until $|Z|\sim20$ kpc
but there is a hint of a small slope of $-0.015\pm0.003$ with
a coefficient of the Pearson correlation of 0.88 for the high-$\mgfe$ sequence 
at $0<|Z|<8$ kpc, beyond which the number of stars is too small for a good statistic.
For the more metal poor end at $-2.0<\feh<-1.6$ and $-2.6<\feh<-2.0$, Fig.~9 shows
no slope in the $\feh$ versus $|Z|$ diagram for the whole $0<|Z|<20$ kpc range,
but there is a hint of a possible gradient for $0<|Z|<8$ kpc. 
In this comparison,
we use $|Z|$, rather than $R$, in order to
reduce the contribution from the Galactic disk.
We note that the Galactic disk is mainly limited within $|Z| < 5$ kpc,
while the Galactocentric distances ($R$) of disk stars 
beyond solar circle (at 8 kpc) could reach farther than 15 kpc \citep{Blant16}.
Thus, the metallicity gradient of the Galactic halo, if existed, 
would be clearer in the
vertical $|Z|$ direction than the Galactocentric distances $R$.

\begin{figure}
\plotone{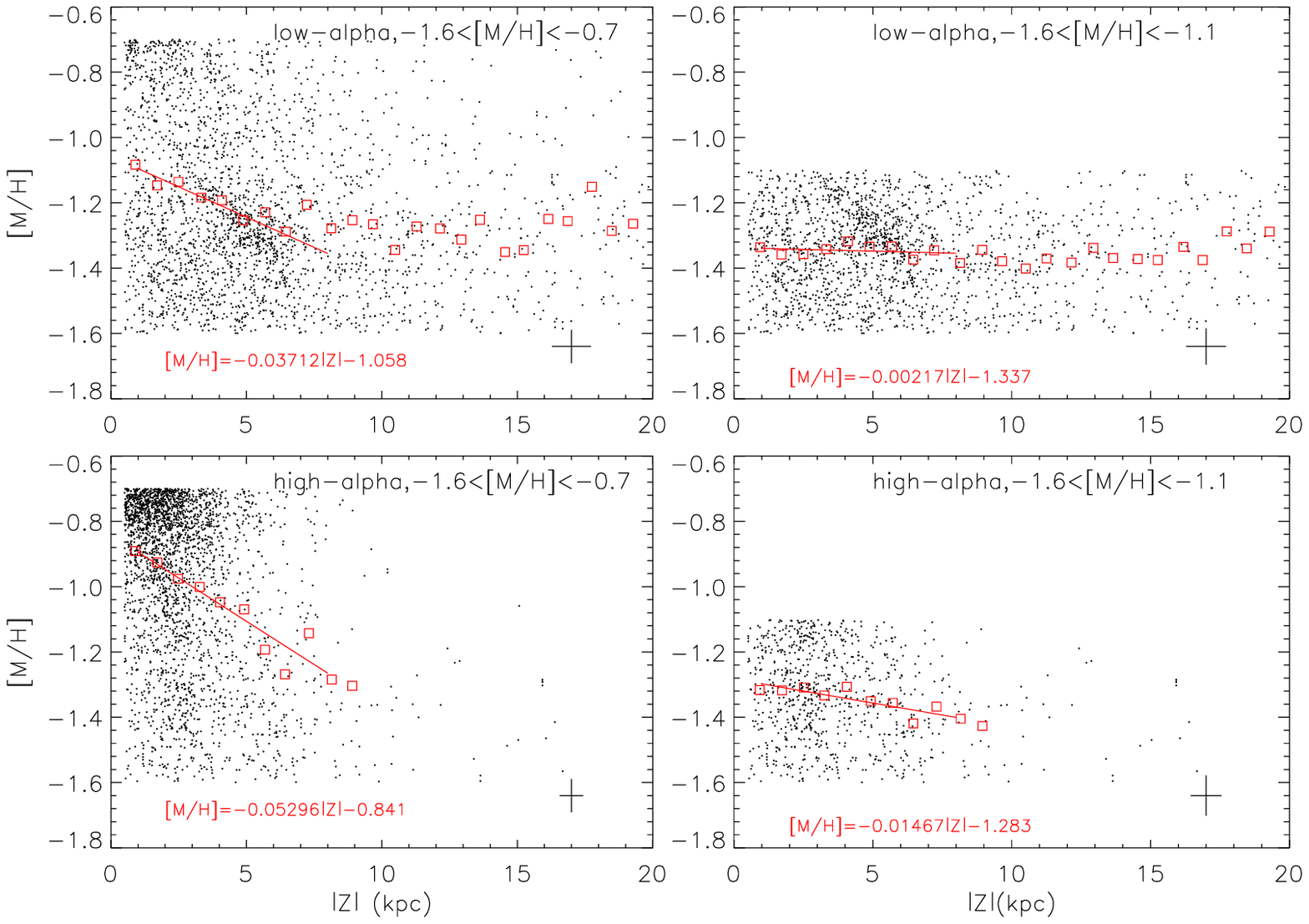}
\caption{Metallicity gradients for low-$\mgfe$ and high-$\mgfe$ sequences
in the two metallicity ranges of $-1.6<\feh<-0.7$ and $-1.6<\feh<-1.1$. 
Open squares are the mean metallicities at given $|Z|$ and solid lines 
are linear regressions to the data at $|Z|<8$ kpc.}
\label{FeZ}
\end{figure}

\begin{figure}
\plotone{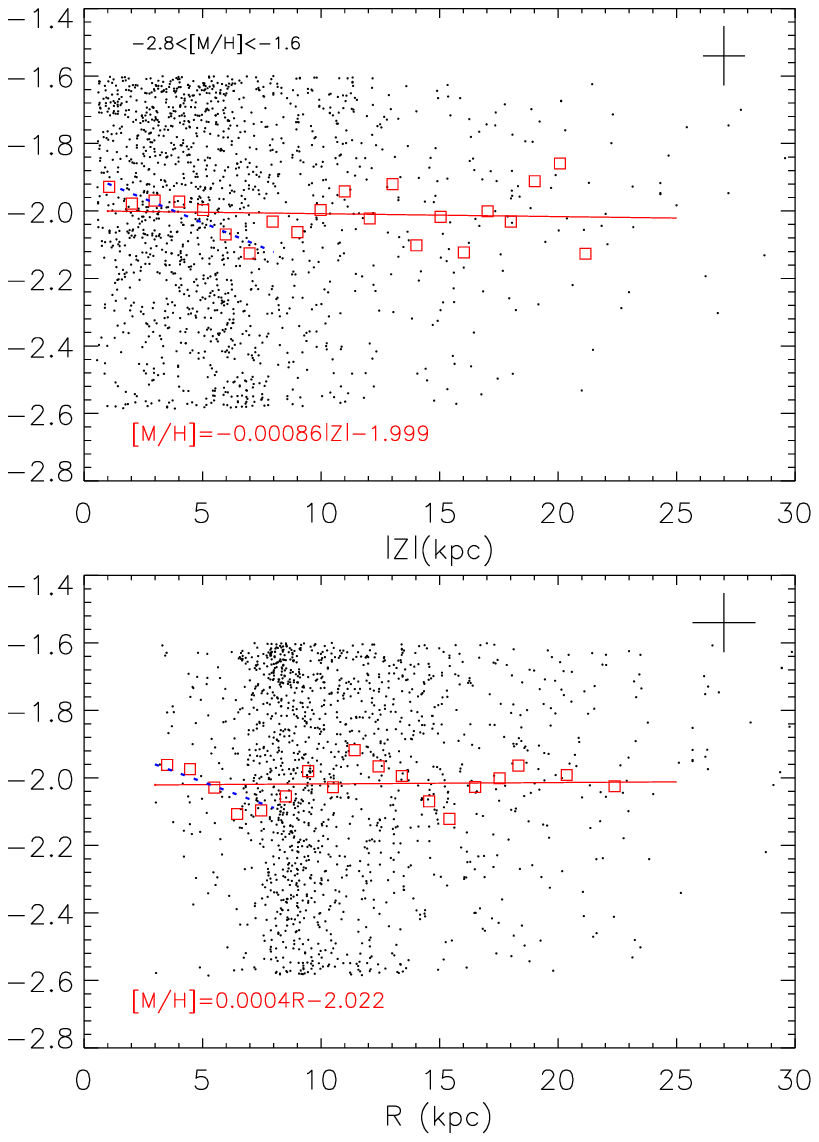}
\caption{The metallicity gradients for metal poor stars
at $-2.0<\feh<-1.6$ and $-2.6<\feh<-2.0$. The symbols are the same as Fig.~8.}
\label{FeZn}
\end{figure}

\section{Implications and Conclusions}
Based on the APOGEE DR14 and the distance catalog, we investigate
the space distributions, $\vlos$ variations along Galactic longitude, 
and abundance gradients between the high-$\mgfe$ and low-$\mgfe$ sequences
in the $\feh$ versus $\mgfe$ diagram. In particular, we investigate 
the distributions of the two $\afe$ sequences far beyond 
the solar vicinity ($D < 5$ kpc) for the first time.

The main results of the present work are summarized as follows.
The two $\mgfe$ sequences are clearly shown for stars
at $5<|Z|<10$ kpc and well separated in the intermediate-metallicity 
range of $-1.1 <\feh <-0.7$. 
In the $\feh$ versus $\mgfe$ diagram,
the transition metallicity
of the halo from the disk is found at $\feh \sim -0.7\pm0.05$ for the low-$\mgfe$ sequence
and of $\feh \sim -1.1\pm0.05$ for the high-$\mgfe$ sequence by investigating 
the $l$ versus $\vlos$ diagrams with different metallicity bins. 
Both the high-$\mgfe$ thick disk with $\feh>-1.1$ and the in situ halo with $-1.6<\feh<-1.1$
have a cutoff at $R\sim 15$ kpc and $ |Z|\sim 8-10$ kpc,
beyond which the low-$\mgfe$ accreted halo with $-1.6<\feh<-0.7$ is dominating.
There is a negative $\mgfe$ gradient for stars at $|Z| > 10$ kpc in the
metallicity range of $-1.1<\feh<-0.7$, but it {flattens out at $|Z|\sim 30$ kpc. 
In the metallicity range of $-2.6<\feh<-1.1$, most stars follow a negative $\mgfe$ gradient
at $|Z|= 10-30$ kpc, but a constant $\mgfe \sim 0.0$ is found for a small fraction of stars
at $|Z|= 10-30$ kpc and it extends outward at $|Z|> 30$ kpc.}
In the intermediate-metallicity range of $-1.6<\feh<-1.1$, the $\feh-|Z|$ gradient 
exists at $|Z|<8$ kpc for the high-$\mgfe$ sequence, while there is no such 
trend for the low-$\mgfe$ sequence.

Based on these results, we may suggest a three-section halo
with different chemical properties:
the inner in situ halo at $|Z|<8-10$ kpc 
with high-$\mgfe$ and a metallicity gradient, 
the intermediately outer halo at $10<|Z|<30$ kpc without metallicity gradient
but with {\bf dual-mode} $\mgfe$ gradient, and the extremely outer accreted halo with $ |Z|>30$ kpc
without both metallicity and $\mgfe$ gradients. 
This suggestion 
is consistent with the transition of the dual-mode halo at $R \sim 15-20$ kpc
and the radius break at $R \sim 30-45$ kpc in the spatial density distribution of halo stars as already described
in the Introduction of the present work. We emphasize that
the selection effect of the APOGEE survey is not considered
in the present paper. But we note that
\cite{Nandakumar17} found 
a negligible selection function effect on the metallicity distribution
function and the vertical
metallicity gradients for the APOGEE survey. In view of this, we expect
that our derived $\mgfe$ and metallicity gradients  may not be affected
by the selection effect.
Further investigations on
more precise observations and theoretical simulations are desirable
to discover the formation of the Galaxy.
In particular, the star number at the distant region of the Galaxy 
is too small presently and we need a deeper spectroscopic survey in the future. 
Meanwhile, the improvement on the distance scale for distant
stars is desirable to ensure that they
have precision as good as that of nearby stars.
Future Thirty Meter Telescope equipped with high-resolution spectrographs
in both near infrared and optical bands is awaiting an opportunity to explore
a high-precision chemical map of the extremely outer halo of the Galaxy.

\acknowledgments
We thank the referee for valuable suggestions and a very careful reading of
the manuscript, which greatly improve the paper. We are grateful to
Prof. Poul Erik Nissen for his constructive comments on the first draft.
We acknowledge our debt to the members of R-software for their continuing 
commitment to the R project.

This study is supported by the National Natural Science
Foundation of China under grant Nos. 11625313, 11890694, 11390371, and 11573035.
Xue X.X. acknowledges the support from the NSFC grant No. 11873052 and
the "Recruitment Program of Global Youth Experts of China".

Funding for the Sloan Digital Sky Survey IV has been provided  by  the  Alfred  P.  Sloan  Foundation,  the  U.S.
Department of Energy Office of Science, and the Participating Institutions.  SDSS-IV acknowledges support and
resources from the Center for High-Performance Computing at the University of Utah.  The SDSS website is www.sdss.org.

SDSS-IV  is  managed  by  the  Astrophysical  Research Consortium  for  the  Participating  Institutions  of  the
SDSS  Collaboration  including  the  Brazilian  Participation  Group,   the  Carnegie  Institution  for  Science,
Carnegie   Mellon   University,   the   Chilean   Participation Group,  the French Participation Group,  Harvard-Smithsonian  Center  for  Astrophysics,  Instituto  de  Astrof\'isica  de  Canarias,  The  Johns  Hopkins  University,
Kavli  Institute  for  the  Physics  and  Mathematics  of the  Universe  (IPMU)/University  of  Tokyo,  Lawrence
Berkeley National Laboratory, Leibniz Institut f\"ur Astrophysik Potsdam (AIP), 
Max-Planck-Institut f\"ur Astronomie  (MPIA  Heidelberg),  Max-Planck-Institut  f\"ur
Astrophysik (MPA Garching), Max-Planck-Institut f\"ur
Extraterrestrische Physik (MPE), National Astronomical Observatories of China,  New Mexico State University,  New  York  University,  University  of  Notre  Dame,
Observatario Nacional/MCTI, The Ohio State University, Pennsylvania State University, Shanghai Astronomical Observatory, United Kingdom Participation Group,
Universidad Nacional Autonoma de Mexico, University of Arizona, University of Colorado Boulder, University
of Oxford, University of Portsmouth, University of Utah,
University of Virginia,  University of Washington,  University  of  Wisconsin,  Vanderbilt  University,  and  Yale
University.

\bibliography{astro_chen.bbl}

\begin{thebibliography}{}
\expandafter\ifx\csname natexlab\endcsname\relax\def\natexlab#1{#1}\fi

\bibitem[{Abadi} {et al.} (2006)]{Abadi06} {Abadi}, M.G., {Navarro}, J.F., {Steinmetz}, M., 2006, MN, 365, 747
\bibitem[{Abolfathi} {et al.} (2018)]{Abolfathi18} {Abolfathi}, B., {Aguado}, D. S., {Aguilar}, G., {et al}. 2018, ApJS, 235, 42
\bibitem[{Akhter} {et al.} (2012)]{Akhter12}{Akhter}, S., {Da Costa}, G.S., {Keller}, S.C., {Schmidt}, B.P.,2012,ApJ,756,23
\bibitem[{Babusiaux} { et al.} (2018)]{Babusiaux18}{Gaia Collaboration},{Babusiaux}, C., {van Leeuwen}, F.,; {Barstow}, M. A., {et~al.} 2018, A\&A, 616, A10.
\bibitem[{Bland-Hawthorn} \& {Gerhard} (2016)]{Blant16} {Bland-Hawthorn}, J., \& {Gerhard} O., 2016, ARA\&A, 54, 529
\bibitem[{Blanton} {et al.} (2017)]{Blanton17}{Blanton}, M.R., {Bershady}, M.A., {Abolfathi}, B. et al. 2017, AJ, 154, 28
\bibitem[{Bressan} {et al.} (2012)]{Bressan12} {Bressan}, A., {Marigo}, P., {Girardi}, L. et al. 2012, MNRAS, 427, 127
\bibitem[{Bullock} \& {Johnston} (2005)]{Bullock05} {Bullock}, J. S., \& {Johnston}, K. V. 2005, ApJ, 635, 931
\bibitem[{Carollo} { et al.} (2010)]{Carollo10}{Carollo}, D., {Beers}, T. C., {Chiba}, M., et al. 2010, ApJ, 712, 692
\bibitem[{Chen} {et al.} (2014)]{Chen14} {Chen}, Y. Q., {Zhao}, G., {Carrell}, K., et al. 2014, ApJ, 795, 52
\bibitem[{Dinescu}, {Girard} \& {van Altena} {} (1999)]{Dinescu99}{Dinescu}, D.I., {Girard}, T.M. \& {van Altena}, W.F. 1999, AJ, 117, 1792
\bibitem[{Eggen}, {Lynden-Bell} \& {Sandage} (1962)]{ELS62} {Eggen} O. J., {Lynden-Bell} D., {Sandage} A. R., 1962, ApJ, 136, 748
\bibitem[{Font} {et al.} (2011)]{Font11}{Font}, A. S., {McCarthy}, I. G., {Crain}, R. A., et al. 2011, MNRAS, 416, 2802
\bibitem[{Fernandez-Alvar} {et al.} (2015)]{Fernandez15}{Fernandez-Alvar}, E., {Allende Prieto}, C., {Schlesinger}, K. J., et al. 2015, A\&A, 577, 81
\bibitem[{Fernandez-Alvar} {et al.} (2018)]{Fernandez18a}{Fernandez-Alvar}, E., {Tissera}, P.B. {Carigi}, L. et al. 2018, astro-ph/1809.02368
\bibitem[{Garc\'ia} {et al.} (2016)]{Garcia16} {Garc\"ia Perez}, A. E., {Allende Prieto}, C., {Holtzman}, J. A., et al.  2016, AJ, 151, 144
\bibitem[{Gunn} {et al.} (2006)]{Gunn06}{Gunn}, J. E., {Siegmund}, W. A., {Mannery}, E. J., et al. 2006, AJ, 131, 2332
\bibitem[{Hayes} {et al.} (2018)]{Hayes18} {Hayes}, C.R., {Majewski}, S.R., {Hasselquist}, S. et al. 2018, ApJ, 852, 49
\bibitem[{Haywood} {et al.} (2018)]{Haywood18}  {Haywood}, M., {Di Matteo}, P., {Lehnert}, M., {Snaith}, O., {Khoperskov}, S., {G\"omez}, A., 2018, ApJ, 863, 113
\bibitem[{Hawkins} {et al.} (2015)]{Hawkins15} {Hawkins}, K., {Jofre}, P., {Masseron}, T., \& {Gilmore}, G. 2015, MNRAS, 453, 758
\bibitem[{Helmi}  {et al.} (2018)]{Helmi18} {Helmi}, A., {Babusiaux}, C., {Koppelman}, H.H., {Massari}, D., {Veljanoski}, J., {Brown}, G. A., 2018, Nature, 563, 85
\bibitem[{Holtzman} {et al.} (2018)]{Holtzman18} {Holtzman}, J.A., {Hasselquist}, S., {Shetrone}, M., et al. 2018, AJ, 156, 125 
\bibitem[{Kellerr} {et al.} (2008)]{Keller08}{Keller}, S.C., { Murphy}, S., {Prior}, S. et al. 2008, ApJ, 678, 851
\bibitem[{Majewski} {et al.} (2017)]{Majewski17} {Majewski}, S. R., {Schiavon}, R. P., {Frinchaboy}, P. M., et al. 2017, AJ, 154, 94
\bibitem[{Myeong} {et al.} (2018)]{Myeong18} {Myeong}, G.C., {Evans}, N.W., {Belokurov}, V., {Sanders}, J.L. \& {Koposov}, S.E. 2018, MNRAS, 478, 5449
\bibitem[{McCarthy} {et al.} (2012)]{McCarthy12} {McCarthy}, I. G., {Font}, A. S., {Crain}, R. A., et al. 2012, MNRAS, 420, 2245
\bibitem[{Nandakumar} {et al.} (2017)]{Nandakumar17} {Nandakumar}, G., {Schultheis}, M., {Hayden}, M. et al. 2017, A\&A, 606, 97
\bibitem[{Nissen} \& {Schuster} (2010)]{NS10} {Nissen}, P.E. \& {Schuster}, J.W. 2010, A\&A, 110, 666
\bibitem[{Queiroz} {et al.} (2018)]{Queiroz18}, A. B. A., {Anders}, F., {Santiago}, B. X et al., 2018, MNRAS, 476, 2556
\bibitem[{Santiago} {et al.} (2016)]{Santiago16}{Santiago}, B.X., {Brauer}, D.E., {Anders}, F., 2016, A\&A, 585, 42
\bibitem[{Searle} \& {Zinn} (1978)]{SZ78}{Searle} L.,\& {Zinn R.}, 1978, ApJ, 225, 357
\bibitem[{Tolstoy} {et al.} (2009)]{Tolstoy09} {Tolstoy}, E., {Hill}, V.,; {Tosi}, M., 2009, ARA\&A, 47, 371
\bibitem[{Wang} {et al.} (2016)]{Wang16} {Wang}, J.L.l, {Shi}, J.R., {Pan}, K.K. et al. 2016, MNRAS, 460, 3179
\bibitem[{Wilson} {et al.} (2010)]{Wilson10} {Wilson}, J. C., {Hearty}, F., {Skrutskie},M. F., et al. 2010, Proc. SPIE, 7735, 77351
\bibitem[{Xue}  {et al.} (2008)]{Xue08} {Xue}, X.X. et al. 2008, ApJ, 684, 1143
\bibitem[Zolotov et al. (2009)]{Zolotov09}Zolotov A., Willman B., Brooks A. M., Governato F., Brook C. B., Hogg D. W., Quinn T., Stinson G., 2009, ApJ, 702, 1058
\end{thebibliography}

\end{document}